\documentclass[aps, prc, reprint, amsmath, groupedaddress, nofootinbib]{revtex4-1}

\usepackage[utf8]{inputenc}
\usepackage{hyperref}
\usepackage{amsmath}
\usepackage{amssymb}
\usepackage{amsfonts}
\usepackage{tabularx}
\usepackage{booktabs}
\usepackage{graphicx}
\usepackage{color}
\usepackage{multirow}
\usepackage[inline]{enumitem}
\graphicspath{{fig/}}
\definecolor{theblue}{RGB}{0,50,230}

\usepackage{hyperref}
\hypersetup{
  colorlinks=true,
  linkcolor=theblue,
  citecolor=theblue,
  urlcolor=theblue
}

\newcommand{\trento}{T\raisebox{-0.5ex}{R}ENTo}

\newcommand{\sqrts}{\sqrt{s_{NN}}}
\newcommand{\T}{\tilde{T}}
\newcommand{\paddedhline}{\noalign{\smallskip}\hline\noalign{\smallskip}}
\newcommand{\dnchdy}{dN_\text{ch}/d\eta}
\newcommand{\dndypP}{dN_\text{pPb}/d\eta}
\newcommand{\dndyPP}{dN_\text{PbPb}/d\eta}
\newcommand{\x}{\mathbf x}
\newcommand{\y}{\mathbf y}
\newcommand{\z}{\mathbf z}
\newcommand{\trans}{^\intercal}

\begin{document}

\title{Constraints on rapidity-dependent initial conditions from\\ charged particle pseudorapidity densities and two-particle correlations}
\author{Weiyao Ke}
\author{J.\ Scott Moreland}
\author{Jonah E.\ Bernhard}
\author{Steffen A.\ Bass}
\affiliation{Department of Physics, Duke University, Durham, NC 27708-0305}
\date{\today}

\begin{abstract}
We study the initial three-dimensional spatial configuration of the quark-gluon plasma produced in relativistic heavy-ion collisions using centrality and pseudorapidity-dependent measurements of charged particle pseudorapidity densities and two-particle pseudorapidity correlations.
A cumulant-generating function is used to parametrize the rapidity dependence of local entropy deposition and to extend arbitrary boost-invariant initial conditions to nonzero beam rapidities.
The model is compared to p+Pb and Pb+Pb charged particle distributions and two-particle correlations, and systematically optimized using Bayesian parameter estimation to extract high-probability initial condition parameters.
The optimized initial conditions are then compared to a number of experimental observables including the pseudorapidity-dependent anisotropic flows, event-plane decorrelations, and flow correlations.
\end{abstract}

\maketitle

\section{Introduction}

High energy nuclear collisions, designed to compress and heat nuclear matter beyond the QCD transition temperature, evidence a strongly coupled QGP liquid that is well described by hydrodynamic simulations with a small but nonzero specific shear viscosity \cite{Muronga:2004sf, Chaudhuri:2006jd, Romatschke:2007mq, Dusling:2007gi, Song:2007ux, Luzum:2008cw}.
These simulations necessitate, among other ingredients, initial conditions for the spatial distribution of energy (or entropy) deposited in the collision.
Although theoretical and phenomenological models have made considerable progress describing aspects of the initial conditions---for example, charged particle yields and multiparticle correlations at midrapidity \cite{Gale:2012rq, Niemi:2015qia, Bhalerao:2015iya, Bernhard:2016tnd}---the search for a comprehensive theory which describes the full three-dimensional structure of the produced fireball remains an outstanding challenge.

Most current state-of-the-art models neglect longitudinal structure in the QGP medium and assume boost invariance \cite{Miller:2007ri, Drescher:2006ca, Schenke:2012wb, Niemi:2015qia, Moreland:2014oya, Chatterjee:2015aja}.
While this is a good approximation at midrapidity on an event-averaged basis in symmetric heavy-ion collisions, it is not clear how the inclusion of local event-by-event longitudinal fluctuations affects the collision dynamics.
Meanwhile, recent interest in the collective phenomena of small asymmetric collisions \cite{Abelev:2014mda, Aad:2014lta, Aad:2013fja, CMS:2012qk, Chatrchyan:2013nka, Khachatryan:2015waa, Khachatryan:2015oea, Khachatryan:2016ibd, Adare:2014keg, Adare:2015cpn, Adare:2015ctn}, where boost-invariant approximations are notably poor, demand a more realistic description of the QGP rapidity profile.
It has also been realized that rapidity dependent model observables provide new sensitive handles on the temperature dependence of the QGP shear viscosity as well as different schemes of longitudinal entropy deposition \cite{Denicol:2015bnf, Bozek:2010bi}.

Rapidity dependent initial condition models have been proposed in a number of previous works, including Monte Carlo Glauber extensions \cite{Bozek:2015bha, Rybczynski:2013yba}, models based on concepts of minijet production, extended strings, flux-tubes and microscopic hadron transport \cite{Wang:1991hta, Zhang:1999bd, Werner:2010aa, Bozek:2015bna, Petersen:2008dd}, as well as recent developments in color-glass condensate (CGC) effective field theory \cite{Schenke:2016ksl, Dumitru:2011wq, Hirano:2012kj}.
Although ideally, one seeks to compare basic model predictions, e.g.\ charged particle yields and flow harmonics as a function of both centrality and rapidity across various collision systems, validating individual models against the full range of measured observables is naturally more complicated than with boost-invariant initial conditions and current efforts to assess model predictions have been limited in scope.

A complementary approach to model-by-model assessment is to parametrize the QGP initial conditions, embed the resulting profiles in realistic transport simulations, and constrain their functional form using a systematic model-to-data comparison \cite{Novak:2013bqa, Pratt:2015zsa, Sangaline:2015isa, Bernhard:2016tnd}.
While such data-driven methods cannot, for example, explain the physical processes which deposit entropy in the collision, they can determine the necessary outcome of initial condition models, provide guidance to first-principle calculations, and be used to robustly quantify QGP transport properties in the presence of confounding model uncertainties.

Constraining parametric initial condition models presents its own unique challenges.
Initial condition calculations feed into hydrodynamic simulations which introduce additional unconstrained model parameters, and evaluating the resulting multistage model using a single set of parameters may involve $\mathcal{O}(10^4)$ hydrodynamic events and significant computational effort.
The effect of each model parameter on final state observables is, in general, highly correlated, and thus model parameters cannot be tuned individually.
Any effort to constrain model parameters, either by brute force methods or by hand, thus quickly become intractable.

Bayesian methods may be used to constrain computationally intensive models which depend on multiple highly correlated parameters \cite{OHagan:2006ba, Higdon:2008cmc, Higdon:2014tva, Dave:pca}.
Recently, the framework was applied to event-by-event boost-invariant viscous hydrodynamics to estimate QGP initial condition and medium properties with quantitative uncertainty \cite{Bernhard:2016tnd}.
In this study, we relax the boost-invariant approximation and perform the first Bayesian analysis of parametric QGP initial conditions including longitudinal structure.

We parametrize salient features of local rapidity-dependent entropy deposition using a cumulant-generating function with tunable mean, standard deviation, and skewness.
Each cumulant of the longitudinal rapidity profile is expressed as a parametric function of participant nucleon densities, and the resulting profiles are normalized to recover established scaling behavior at midrapidity \cite{Moreland:2014oya, Bernhard:2016tnd}.
The approach is used to investigate two simple models for asymmetric entropy deposition: one where the skewness parameter is proportional to the relative difference in participant nucleon density of each nucleus, and a second proportional to the absolute difference.
These models are chosen to bracket a reasonable range of underlying scaling behavior and can be used to estimate model uncertainties in the unknown form of the chosen parametization.

The global parameter estimation methods employed in this work necessitate a fast and efficient method to calculate the predictions of the model using an arbitrary set of parameter values.
For this purpose, we evaluate a comprehensive hybrid model that couples viscous hydrodynamics to a hadronic afterburner using a relatively small number of parameter configurations and interpolate the result using a Gaussian process emulator \cite{Rasmussen:2006gp}.
While this works well for boost-invariant simulations, 3+1D viscous hydrodynamics requires an order-of-magnitude more computation time.
We mitigate this additional overhead by running our 3+1D hydrodynamic simulations \cite{Karpenko:2013wva} in ideal (nonviscous) mode on a coarse space-time grid which significantly accelerates the computation.
We argue that the use of ideal hydrodynamics is acceptable since we calibrate the model on multiplicity observables which receive little modification from QGP viscosity.
It has also been shown that the two particle pseudorapidity correlations are largely insensitive to viscosity of the QGP phase \cite{Denicol:2015bnf}, although we do include a UrQMD hadronic afterburner in the calculation of these observables \cite{Bass:1998ca, Bleicher:1999xi}.

Once the parametric initial conditions have been constrained to fit p+Pb and Pb+Pb charged particle pseudorapidity densities and pseudorapidity correlations, we use high likelihood initial condition parameter sets with a 3+1D viscous hybrid model \cite{Karpenko:2013wva} to calculate novel three-dimensional observables such as pseudorapidity-dependent anisotropic flows, event-plane decorrelations and flow correlations.
We find that both parametrizations simultaneously describe p+Pb and Pb+Pb charged particle distributions and central to mid-central pseudorapidity correlations in Pb+Pb collisions quite well.

\section{Three-dimensional Initial Condition Model}

In the present study we seek to parametrize and constrain the full three-dimensional structure of the produced QGP medium.
It is thus essential that any resulting model maintain good agreement with charged particle multiplicity distributions and flow harmonics at midrapidity.
We consequently adopt a factorized approach and express the initial entropy density $s$ at the hydrodynamic starting time $\tau_0$ as
\begin{equation}
  s(\x, \eta_s)\vert_{\tau=\tau_0} \propto f(\x) \times g(\x, \eta_s),
  \label{factorized}
\end{equation}
where $\x = (x, y)$ is a position in the transverse plane, $\eta_s$ is the spacetime rapidity, $f$ denotes the entropy density in the transverse plane at midrapidity, and $g$ is some rapidity-dependent function such that $g(\x, 0)=1$.

\subsection{Midrapidity calculation}
We simulate entropy deposition at midrapidity using the parametric model \trento\ \cite{Moreland:2014oya}, which is constructed to interpolate a subspace of all initial condition models including (but not limited to) specific calculations in CGC effective field theory.
The initial condition model was recently embedded in an event-by-event hybrid model which couples viscous hydrodynamics to a microscopic kinetic description \cite{Shen:2014vra} and its parameters constrained using Bayesian parameter estimation \cite{Bernhard:2016tnd}.
Here we briefly summarize its key features.

The model samples independent nucleon positions from spherical (or deformed) Woods-Saxon density distributions, shifts the nucleons in each nucleus by a random impact parameter offset $b$, and projects each coordinate onto the transverse plane.
Participant nucleons are then sampled according to the pairwise inelastic collision probability
\begin{equation}
  \label{dsigma_db}
  \dfrac{d\sigma_{NN}^\text{inel}}{2\pi b\, db} = 1 - \exp\left[-\sigma_{gg} T_{pp}(b)\right],
\end{equation}
where $b$ is now the impact parameter between two nucleons, $\sigma_{gg}$ is an effective partonic cross section tuned to fit the measured inelastic proton-proton cross section at a given beam energy, and $T_{pp}$ is the proton-proton overlap function
\begin{equation}
  T_{pp}(b) = \int d^2\x\, T_p(\x)\,T_p(\x - \mathbf b).
\end{equation}
The proton thickness function $T_p$ is described by a simple Gaussian density profile
\begin{equation}
  T_p(\x) = \frac{1}{2 \pi w^2} \exp\bigg({-}\frac{\x^2}{2 w^2}\bigg),
  \label{nucleon_profile}
\end{equation}
with effective proton width $w$.
Once the participants in each nucleus are determined according to Eq.~\eqref{dsigma_db}, a participant nuclear thickness function $\T$ is constructed by summing the proton thickness function of each wounded nucleon, e.g., the participant thickness function of nucleus $A$ is given by
\begin{equation}
  \T_A(\x) = \sum\limits_{i=1}^{N_\text{part}} w_i\, T_p(\x - \x_i).
\end{equation}
Here $w_i$ is a random weight sampled from a gamma distribution with unit mean and variance $1/k$, where $k$ is a tunable shape parameter.
This additional source of fluctuation is added to reproduce the negative binomial distribution of the charged particle multiplicity in minimum bias proton-proton collisions \cite{Bozek:2013uha}.

Initial entropy deposition at midrapidity is then described by an eikonal function $s(\x, 0) \propto f(\x)$ which maps participant nucleon density to local entropy deposition.
For this mapping, the \trento\ model adopts a flexible parametric form known as the generalized mean,
\begin{equation}
  \label{gen_mean}
        f(\x) \propto \left( \frac{\T_A^p + \T_B^p}{2} \right)^{1/p},
\end{equation}
where the continuous parameter $p$ smoothly interpolates among different types of entropy deposition schemes \cite{Bernhard:2016tnd}.
For example, $p=1$ is exactly a wounded nucleon model, while $p\sim-0.67$ simulates entropy deposition in the original KLN model \cite{Drescher:2006ca}, and $p \sim 0$ closely mimics the behavior of both IP-Glasma \cite{Schenke:2012wb} and EKRT \cite{Niemi:2015qia} saturation calculations.

\trento\ initial condition parameters have been constrained using Bayesian parameter estimation and calibrated to fit identified particle yields, mean transverse momenta, and flow cumulants in ${\sqrts=2.76}$~TeV Pb+Pb collisions \cite{Bernhard:2016tnd}.
The analysis established 90\% credible intervals for the effective nucleon width {$w \approx 0.5 \pm 0.1$ fm} and entropy deposition parameter ${p \approx 0.0 \pm 0.2}$, while the nucleon fluctuation shape parameter $k$ was essentially unconstrained by data.
Corresponding model predictions simultaneously fit charged particle yields, mean transverse momenta, and flow cumulants at the 10\% level or better across all centralities and hence corroborate the effective initial condition mappings predicted by IP-Glasma and EKRT theory calculations.

\subsection{Extension to nonzero rapidity}

The factorized expression in Eq.~\eqref{factorized} extends entropy deposition at midrapidity $f(\x)$ to nonzero rapidity using a rapidity-dependent mapping $g(\x, \eta)$ which multiplies $f$ to incorporate nontrivial longitudinal structure.
Before constructing the form of this longitudinal dependence, we elucidate our use of spacetime rapidity $\eta_s$, rapidity $y$, and pseudorapidity $\eta$.

Assuming for simplicity that, at early times, the initially produced partons are massless and free-streaming in the $z$ direction, then
\begin{equation}
  \frac{z}{t} = \frac{p_z}{|\mathbf p|}.
\end{equation}
This approximation allows the equivalence of $\eta_s$ and $\eta$ at early stages of the collision:
\begin{equation}
  \eta_s = \frac{1}{2}\log\frac{t+z}{t-z} \sim \eta = \frac{1}{2}\log\frac{|\mathbf p|+p_z}{|\mathbf p|-p_z}.
\end{equation}
Experimentally, the event-averaged rapidity $(y)$ distribution of charged particles in proton-proton collisions resembles a Gaussian, while the conversion to pseudorapidity creates a dip at midrapidity due to the Jacobian $dy/d\eta$.
Therefore, we first parametrize the rapidity dependence of the system and then perform a change of variables from $y$ to $\eta$ using the relations
\begin{align}
  g(\x, \eta)\, d\eta &= g(\x, y)\, dy, \\[1ex]
  \frac{dy}{d\eta} &= \frac{J \cosh \eta}{\sqrt{1 + J^2 \sinh^2 \eta}},
  \label{jacobian}
\end{align}
where the species-dependent factor $J$ is replaced with an effective value $J \approx \langle p_T \rangle / \langle m_T \rangle$.
We then invoke $\eta_s \sim \eta$ for an initial condition of massless partons, so that the rapidity-dependent entropy profile is
\begin{equation}
  s(\x, \eta_s)\vert_{\tau=\tau_0} \propto f(\x)\, g(\x, y)\, \frac{dy}{d\eta}.
\end{equation}

\begin{table}
  \caption{
    \label{tab:parametrization}
    Rapidity-dependent initial condition parametrizations with two different models for the skewness parameter. The constant $T_0 = 1$~fm$^{-2}$ preserves desired dimensionality.
  }
  \begin{ruledtabular}
    \begin{tabular}{lccc}
      & \multicolumn{3}{c}{Distribution cumulant:} \\
      \noalign{\smallskip}\cline{2-4}\noalign{\smallskip}
      Model & \multicolumn{1}{c}{mean $\mu$} & \multicolumn{1}{c}{std.\ $\sigma$} & \multicolumn{1}{c}{skewness $\gamma$} \\
      \paddedhline \noalign{\smallskip}
        Relative  & $\frac{1}{2} \mu_0 \log(T_A/T_B)$ & $\sigma_0$ & $\gamma_0 \dfrac{T_A - T_B}{T_A + T_B}$ \smallskip\\
        Absolute & $\frac{1}{2} \mu_0 \log(T_A/T_B)$  & $\sigma_0$ & $\gamma_0 (T_A - T_B)/T_0$\smallskip
    \end{tabular}
  \end{ruledtabular}
\end{table}

At this point, we require a parametric mapping $g(\x, y)$ which encodes nontrivial longitudinal structure and extends the model to forward and backward rapidity.
Rather than assert an explicit functional form, we parametrize $g$ using cumulants and construct the function from the inverse Fourier transform of its cumulant-generating function,
\begin{align}
  g(\x, y) &= \mathcal{F}^{-1}\{\tilde{g}(\x, k)\}, \label{gfunc} \\
  \log \tilde{g} &=  i \mu k - \frac{1}{2}\sigma^2 k^2 - \frac{1}{6} i \gamma \sigma^3 k^3 + \cdots \label{cgf}
\end{align}
The function is then normalized, $g(\x, 0)=1$, in order to preserve the desired scaling behavior at midrapidity.
This ansatz naturally enables us to control different aspects of the longitudinal entropy deposition (width, skewness, etc) independently.
In the present study, we consider the first three cumulants; including higher-order cumulants is possible but increases the model complexity.

Different rapidity-dependent initial condition models are described by different parametrizations of the generating-function cumulants.
Two existing approaches include so-called ``shifted" and ``tilted" models for longitudinal entropy deposition \cite{Bozek:2010bi}.
Shifted initial conditions assume that each participant's entropy profile is Gaussian in rapidity space with its mean rapidity shifted to the center of mass rapidity $\eta_\text{cm}=\frac{1}{2} \log (T_A/T_B)$.
Alternatively, tilted initial conditions omit a translational rapidity shift and opt for a linear tilting factor
\begin{align}
  s(\x, \eta_s) = s(\x) [1+\eta_s\, \mathcal{A}(\x)],
\end{align}
where $\mathcal{A}$ is some local measure of nuclear thickness asymmetry, with the property that
\begin{align}\label{asym}
  \mathcal{A}(T_A, T_B) = -\mathcal{A}(T_B, T_A).
\end{align}
In terms of cumulants, the shifted model alters the mean $\mu$ of the local rapidity distribution, and the tilted model mainly affects the skewness $\gamma$.
However, in general, all of the first few cumulants of the distribution could be nonzero, i.e.\ nature may opt for an initial entropy deposition scheme which is both shifted and tilted along the beam axis.

We therefore parametrize the first three cumulants $\mu$, $\sigma$, and $\gamma$ of the local rapidity distribution $g(\x, y)$ using three corresponding coefficients $\mu_0$, $\sigma_0$, and $\gamma_0$ which modulate the local rapidity distribution's shift, width, and skewness, respectively.
These parametrizations, listed in Table~\ref{tab:parametrization}, include two different models for the skewness factor $\gamma$.
The first, a relative-skewness model, uses a common dimensionless asymmetry measure
\begin{equation}
  \mathcal{A}(T_A, T_B) = \gamma_0\frac{T_A - T_B}{T_A + T_B},
\end{equation}
which saturates when $T_A \gg T_B$ and vice versa, while the second, an absolute-skewness model, employs a dimensionful construction given by
\begin{equation}
  \mathcal{A}(T_A, T_B) = \gamma_0 \frac{T_A - T_B}{T_0},
\end{equation}
where $T_0=1$~fm$^{-2}$ restores the desired dimensionality.

\begin{figure}[t]
  \includegraphics{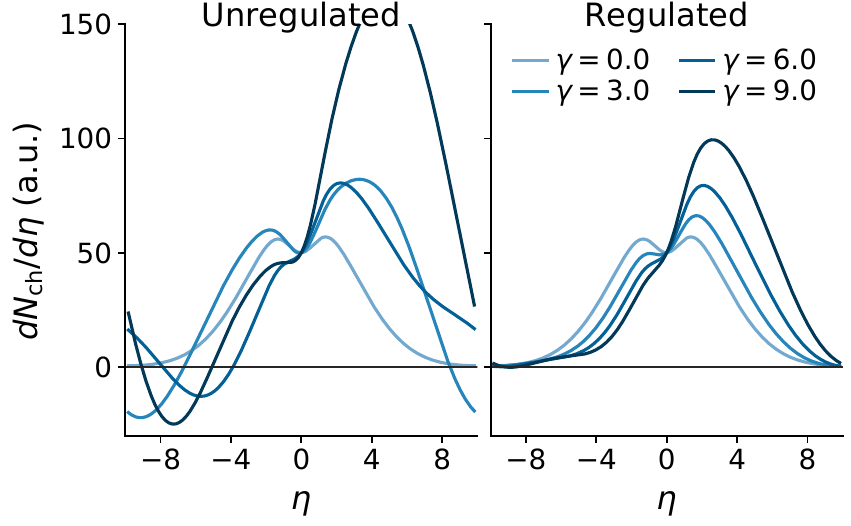}
  \caption{Left: Unregulated skewness values lead to ill-behaved rapidity distributions. The distributions scale non-monotonically with skewness parameter $\gamma$ and go negative at large rapidities. Right: Replacing $\gamma$ by Eq.~\eqref{regulateEq} achieves desired monotonic scaling and suppresses negative regions.}
  \label{fig:regulate}
\end{figure}

\begin{figure}[b]
  \includegraphics{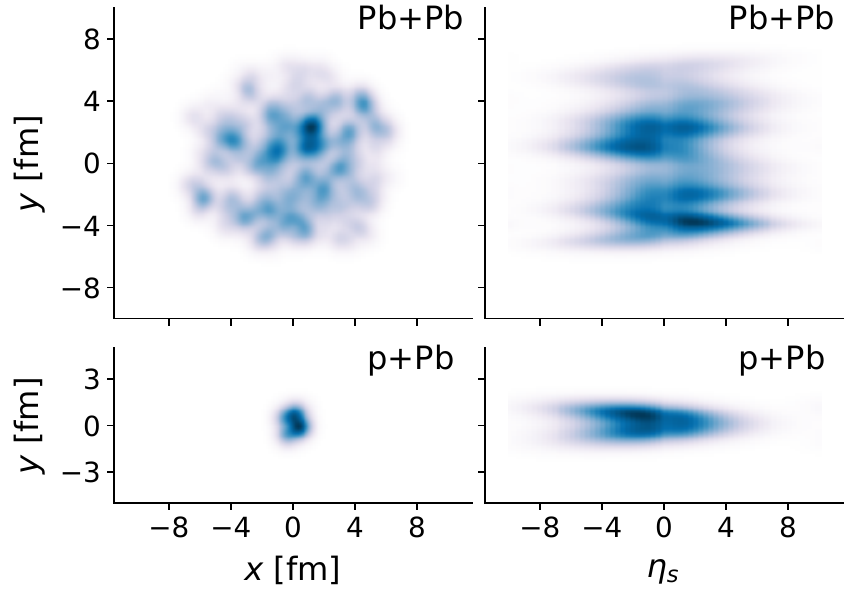}
  \caption{Initial entropy density in sample Pb+Pb (upper) and p+Pb (lower) events for cross sections of the $\eta=0$ and $x=0$ planes (left and right columns). Event is constructed using the relative skewness model in Table~\ref{tab:parametrization} with $\mu_0=1$, $\sigma_0=3$ and $\gamma_0=6$ along with midrapidity parameters from \cite{Bernhard:2016tnd}.}
  \label{fig:3d-example}
\end{figure}

There is, however, a problem with the reconstructed function in Eq.~\eqref{gfunc}.
Rapidity distributions with large skewness are ill-behaved and oscillate to negative values for large $\eta$, as shown in the left panel of Fig.~\ref{fig:regulate}.
The conditions to ensure a positive definite Fourier transform are quite involved; instead, we employ the following substitution for $\gamma$ to regulate the distribution:
\begin{equation}
  \gamma \rightarrow \gamma \exp \left({-}\frac{1}{2}\sigma^2k^2 \right). \label{regulateEq}
\end{equation}
This leaves the skewness of the distribution intact while contributions from higher-order cumulants are included systematically.
The performance of the procedure depends on the domain of reconstruction.
For reasonable values of skewness, it is found to perform well within the range ${-3.3\, \sigma \le \eta \le 3.3\, \sigma}$.
Fig.~\ref{fig:regulate} shows the reconstructed function before and after the regulation procedure using four different skewness values.
The procedure strongly suppresses negative regions of charged particle density (in realistic calculations they are set to zero), and preserves a clear monotonic trend with increasing skewness.
The regulated generating function approach thus produces distributions which are both positive and well behaved over a wide range of rapidity and skewness.

In Fig.~\ref{fig:3d-example} we show the resulting entropy density $s(\x, \eta_s)$ produced by randomly generated Pb+Pb and p+Pb collisions (top and bottom panels respectively) using a set of typical parameter values, annotated in the figure caption.
The left column shows the entropy density at midrapidity $\eta_s=0$, while the right column shows a cross section of the entropy density in the plane defined by $x=0$.
We observe large entropy density fluctuations in the transverse plane arising from nucleon position fluctuations as well as significant forward-backward rapidity fluctuations which track the local asymmetry of projectile and target nucleon densities.
The event profiles---although not yet optimized---exhibit rich longitudinal structures which clearly break boost invariance.

At this point it is important to emphasize that the present entropy deposition parametrization is a purely \emph{local function} of nuclear overlap density.
In subsequent sections, we optimize the model parameters to fit \emph{global} pseudorapidity-dependent charged particle yields which have their transverse structure integrated out.
Obtaining quality fits to data thus remains a nontrivial task, as the model convolves a local entropy deposition mapping with a background distribution of fluctuating nuclear overlap density.

\section{Experimental observables}
The three-dimensional initial entropy profile and its longitudinal fluctuations directly relate to the average charged particle multiplicity and event-by-event fluctuations observed in the final state. 
The ALICE collaboration has published the centrality dependent pseudorapidity densities $\dnchdy$ in Pb+Pb collisions with a wide pseudorapidity coverage $-3.5<\eta<5.0$ \cite{Abbas:2013bpa,ALICE:2015kda}, and the ATLAS collaboration has measured centrality-dependent $\dnchdy$ in p+Pb collisions within $|\eta| < 2.7$ \cite{Aad:2015zza}.
However, such centrality binned $\dnchdy$ measurements only probe the initial entropy density averaged over many events. 
Information on the event-by-event fluctuations is carried by two-particle pseudorapidity correlation $C(\eta_1, \eta_2)$.
In particular, the long range contributions to the correlation function (LRC) are sensitive to the event-by-event initial state fluctuations and are ideal for examining stochastic properties of the three-dimensional initial conditions.

We follow \cite{Bzdak:2012tp, Jia:2015jga, ATLAS:2015kla} and decompose the event-by-event charged particle pseudorapidity distribution within the acceptance $[-Y, Y]$ using normalized Legendre polynomials
\begin{align}
  \frac{dN}{d\eta} &= \biggl\langle\frac{dN}{d\eta}\biggr\rangle \biggl[1 + \sum_{n=0}^\infty a_n T_n\left(\frac{\eta}{Y}\right) \biggr], \\[.5ex]
  T_n(x) &= \sqrt{n + \frac{1}{2}}\, P_n(x).
\end{align}
The two-particle correlation function $C(\eta_1, \eta_2)$ is then calculated from the normalized multiplicity distribution $R(\eta) = dN/d\eta /\langle dN/d\eta\rangle$ and is decomposed into symmetrized polynomials $T_{mn}(\eta_1, \eta_2)$
\begin{align}
  C(\eta_1, \eta_2) &= \left\langle R(\eta_1) R(\eta_2)\right\rangle \\
  &= 1 + \sum_{m, n}\langle a_m a_n\rangle  T_{mn}(\eta_1, \eta_2),  \\
  T_{mn}(\eta_1, \eta_2) &= \frac{T_n(\eta_1)T_m(\eta_2) + T_m(\eta_1)T_n(\eta_2)}{2}.
\end{align}
Centrality fluctuations introduce nonzero $\langle a_0 a_n\rangle$ and are removed by renormalizing $C(\eta_1, \eta_2)$,
\begin{align}
  C_N(\eta_1, \eta_2) &= \frac{C(\eta_1, \eta_2)}{C_1(\eta_1)C_2(\eta_2)},\\[.5ex]
  C_{1,2}(\eta_{1,2}) &= \int_{-Y}^{Y}C(\eta_1, \eta_2)\frac{d\eta_{2,1}}{2Y}.
\end{align}
The forward-backward multiplicity fluctuations characterized by $\langle a_m a_n\rangle$ with $m, n > 0$ can then be projected out from the renormalized correlation function as
\begin{align}
  C_N(\eta_1, \eta_2) \sim 1 + \frac{3}{2}\langle a_1 ^2 \rangle \frac{\eta_1\eta_2}{Y^2} + \cdots.
\end{align}

The ATLAS collaboration has measured the centrality dependence of various $\langle a_m a_n\rangle$ in Pb+Pb collisions \cite{ATLAS:2015kla, SoorajRadhakrishnanfortheATLAS:2015eqq} using particles with $p_T > 0.5$~GeV.
Recent theoretical work on these observables includes a longitudinal extension of IP-Glasma initial conditions \cite{Schenke:2016ksl} and a rapidity-dependent constituent-quark MC-Glauber model which was embedded in three-dimensional hydrodynamic simulations \cite{Denicol:2015bnf, Monnai:2015sca}.
It was shown that short range correlations (SRC) from resonance decays are a significant contribution to the $\langle a_m a_n\rangle$ signal, while variations in the transport coefficients have a much smaller effect \cite{Denicol:2015bnf}.
The contributions from LRC and SRC were estimated in a subsequent ATLAS analysis \cite{Jia:2016jlg} using correlations of same- and opposite-signed charged particles with $p_T > 0.2 \textrm{ GeV}$.
The isolated LRC, which were measured with a lower $p_T$ cut, are a cleaner observable for the study of initial state fluctuations, but they are not yet calculated for central Pb+Pb collisions.
We therefore perform the calculation \cite{ATLAS:2015kla} with proper modeling of the SRC using the UrQMD model.

\section{Model-to-data comparison}

\begin{table}[t]
  \caption{Input parameter ranges for the rapidity-dependent parametric initial condition model.}
  \begin{ruledtabular}
    \begin{tabular}{lll}
      Parameter & Description	& Range \\
      \paddedhline
      $N_{\textrm{p+Pb}}$    & Overall p+Pb normalization      & 140.0--190.0 \\
      $N_{\textrm{Pb+Pb}}$   & Overall Pb+Pb normalization     & 150.0--200.0  \\
      $p^\dagger$	                   & Generalized mean parameter      & -0.3--0.3  \\
      $k$	                   & Multiplicity fluct.\ shape      & 1.0--5.0  \\
      $w$	                   & Gaussian nucleon width     & 0.4--0.6  \\
      $\mu_0$                & Rapidity shift mean coeff.\     & 0.0--1.0  \\
      $\sigma_0$             & Rapidity width std.\ coeff.\    & 2.0--4.0  \\
      \multirow{2}{*}{$\gamma_0$}             & \multirow{2}{*}{Rapidity skewness coeff.\ }      & 0.0--10.0 (rel) \\
                  &        & 0.0--3.6 (abs)  \\
      $J$	                   & Pseudorapidity Jacobian param.  & 0.6--0.9
    \end{tabular}
  \end{ruledtabular}
   \raggedright{$\dagger$ Priori probability distributions fitted from \cite{Bernhard:2016tnd} are applied on this parameter independently within the given ranges.}
  \label{tab:parameters}
\end{table}

The aforementioned rapidity extension introduces several new model parameters which necessitate rigorous optimization.
For this purpose, we apply established Bayesian methodology \cite{OHagan:2006ba, Higdon:2008cmc, Higdon:2014tva, Wesolowski:2015fqa} to constrain the proposed parametric initial conditions and extract intrinsic, local features of the QGP fireball using macroscopic event-averaged quantities.
Ideally, one would run the full model calculation at each design point and calibrate the model to fit a comprehensive list of experimental measurements.

Unfortunately, three-dimensional viscous hydrodynamic simulations require an order of magnitude more computing resources than the boost-invariant models previously used in Bayesian analyses.
This makes it difficult to calibrate on statistically intensive observables such as multiparticle flow correlations which require tens of thousands of minimum bias events at each design point. 

Work is underway to solve these technical challenges by migrating three-dimensional viscous hydrodynamic simulations to graphics cards which offer dramatic performance enhancements over processors at a fraction of the cost \cite{Bazow:2016yra}.
We instead omit higher order flow observables and calibrate only on the rapidity-dependent charged particle yields and two-particle pseudorapidity correlations which can be calculated with a few thousand events. 

It should also be noted that the data we use for Pb+Pb and p+Pb collisions are taken at different beam energies, ${\sqrts = 2.76}$~TeV and 5.02~TeV respectively.
Because phenomenological model parameters generally change with beam energy, it is not fully consistent to optimize a single set of parameters to fit experimental observables at two different energies.
Here we assume that all model parameters, except for the overall entropy normalizations, do not change drastically with the beam energy of the two datasets since the beam rapidity changes less than $8\%$ and perform a simultaneous multi-parameter fit using both measurements.

We now briefly summarize the procedure used to apply Bayesian methodology to the newly constructed parametric initial condition model. For a more comprehensive explanation see \cite{Novak:2013bqa, Bernhard:2015hxa, Bernhard:2016tnd}. All steps are repeated for both the relative and absolute skewness models described in Table~\ref{tab:parametrization}.

\subsection{Parameter design}

The three-dimensional parametric initial conditions are specified using nine model parameters.
Five control entropy deposition at midrapidity:
\begin{enumerate}[itemsep=0pt]
  \item[1--2.] two normalization factors for Pb+Pb and p+Pb collisions at $\sqrts=2.76$~TeV and 5.02~TeV beam energies,
  \item[3.] the generalized mean parameter $p$ which modulates entropy deposition at midrapidity,
  \item[4.] a gamma shape parameter $k$, which controls the variance of proton-proton multiplicity fluctuations,
  \item[5.] and a Gaussian nucleon width $w$, which determines initial state granularity;
\end{enumerate}
the remaining four parameters add rapidity dependence to the model:
\begin{enumerate}[itemsep=0pt]
  \setcounter{enumi}{6}
  \item[6--8.] three coefficients which modulate the local rapidity distribution's shift $\mu_0$, width $\sigma_0$, and skewness $\gamma_0$,
  \item[9.] and a Jacobian factor $J$ for the conversion from rapidity to pseudorapidity.
\end{enumerate}

Given the large number of iterations required by multidimensional Monte Carlo optimization methods and the non-negligible computation time needed to sample initial condition events, it is intractable to explore the aforementioned parameter space using direct model evaluation.
To circumvent this issue, we train emulators using a limited number of parameter configurations to reproduce the charged-particle pseudorapidity density and the two-particle pseudorapidity correlations predicted by the initial condition model.
These emulators interpolate the predictions of the model between training points and provide essentially instant predictions at uncharted regions of parameter space.

The emulators are trained using 100 unique parameter configurations sampled from the parameter ranges listed in Table~\ref{tab:parameters}.
Each parameter design point is distributed in the nine dimensional space using a maximin Latin hypercube design---a space-filling algorithm that maximizes the minimum distance between pairs of points in the multidimensional space.

With the parameter design in hand, we run $4\times 10^3$ Pb+Pb and $10^4$ p+Pb initial condition events through hydrodynamics at each of the 100 points and calculate the charged-particle pseudorapidity density and two-particle pseudorapidity correlations.
The centrality bins are defined by charged particle multiplicity using the same kinematic cuts used by experiments: $|\eta|<0.8$ for Pb+Pb collisions and ${-4.9 < \eta < -3.1}$ for p+Pb collisions.
The resulting $\dnchdy$ and rms $a_1$ are concatenated to an observable array for each input parameter set.
Loosely speaking, the physics model maps the $m\times n$ parameter design matrix $X$ to an $m \times p$ observable matrix $Y$:
\begin{equation}
  \begin{pmatrix}
    x_{1,1} & \cdots & x_{1,n} \\
    \vdots  & \ddots & \vdots \\
    x_{m,1} & \cdots & x_{m,n} \\
  \end{pmatrix}
  \xrightarrow{\text{Model}}
  \begin{pmatrix}
    y_{1,1} & \cdots & y_{1,p} \\
    \vdots  & \ddots & \vdots \\
    y_{m,1} & \cdots & y_{m,p} \\
  \end{pmatrix},
  \label{design-obs}
\end{equation}
where $m=100$ is the number of design points, ${n=9}$ is the number of input parameters, and $p$ is the number of measured outputs.

\subsection{Model emulator}

\begin{figure*}
  \includegraphics{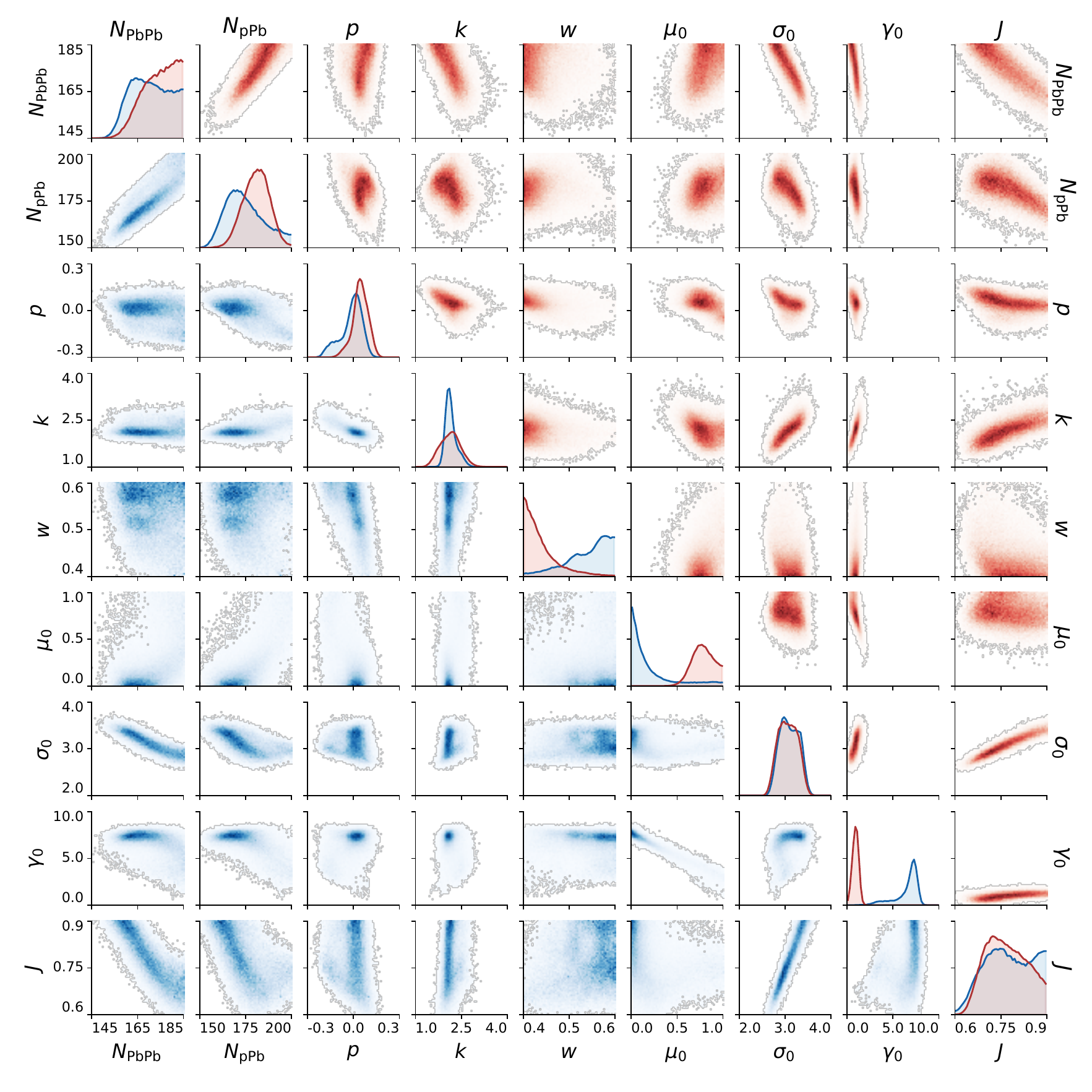}
  \caption{Posterior distributions of the model parameters, listed in Table~\ref{tab:parameters}, for the relative-skewness (blue lower diagonal) and absolute-skewness (red upper diagonal) models. The diagonal panels are the marginal likelihood distributions of individual model parameters, while off-diagonal panels are joint distributions for pairs of model parameters.}
  \label{fig:posterior}
\end{figure*}

Once the design matrix is specified and the model has been run at each design point to construct a corresponding observable matrix, we train a set of Gaussian process emulators to reproduce the predictions of the model.
A Gaussian process emulator is a powerful nonparametric regression method which can be used to interpolate a scalar function of one or more input parameters.
For example, given a set of multivalued inputs ${X=\{\x_1, \x_2, \ldots, \x_n\}}$, and a corresponding list of scalar outputs $\mathbf y=\{y_1, y_2, \ldots, y_n \}$, a Gaussian process emulator can be used to interpolate the function $f: \x \mapsto y$ subject to a pre-specified covariance function $\sigma(\x,\x')$.
We use an existing implementation of Gaussian process regression in this work \cite{2014arXiv1403.6015A}.

Since Gaussian processes are fundamentally scalar functions, while our model produces vector outputs, we first transform the model outputs (normalized by experimental data) using principal component analysis.
The principal components $z$ are orthogonal, uncorrelated linear combinations of the original outputs and may be used to reduce the dimensionality of the output space, since the first several principal components often account for the majority of the model's variance \cite{Dave:pca}.
Independent Gaussian processes are thus trained to emulate the first $q < p$ principal components of the centrality-dependent charged-particle pseudorapidity distribution for the two systems and the rms $a_1$ of Pb+Pb collisions separately.

We choose to include $q=6,$ $6,$ and $4$ principal components for $\dndyPP$,\, $\dndypP$ and rms $a_1$ which account for 99.5\% of the observed variance.
For the emulator covariance function $\sigma(\x, \x')$, we adopt a simple Gaussian form
\begin{equation}
  \sigma(\x, \x') = \sigma_\text{GP}^2 \exp\Biggl[ -\sum_{k=1}^n \frac{(x_k - x'_k)^2}{2\ell_k^2} \Biggr] + \sigma_n^2\delta_{\x\x'},
\end{equation}
which is well-suited for continuously differentiable, smoothly-varying models.
The variance of the Gaussian process $\sigma^2_\text{GP}$, correlation lengths $l_k$, and noise variance $\sigma_n^2$ are estimated from the model outputs by numerically maximizing the likelihood
\begin{equation}
  \log P = -\frac{1}{2}\mathbf y\trans \Sigma^{-1} \mathbf y - \frac{1}{2} \log | \Sigma | - \frac{m}{2} \log 2 \pi,
\end{equation}
where $\Sigma$ is the covariance matrix from applying the covariance function $\sigma$ to each pair of design points.

\subsection{Bayesian calibration and Bayes factor}

The parameter space of the emulated model is finally explored using Markov chain Monte Carlo (MCMC) methods.
The posterior probability for the \emph{true} model parameters $\x_\star$ is given by Bayes' theorem,
\begin{equation}
  P(\x_\star| X, Y, \y_\text{exp}) \propto P(X, Y, \y_\text{exp} | \x_\star) P(\x_\star).
  \label{bayes}
\end{equation}
The left-hand side is the Bayesian \emph{posterior}, the probability of true parameters $\x_\star$ given model design $X$, observable matrix $Y$, and experimental data $\y_\text{exp}$.
On the right, $P(X, Y, \y_\text{exp}| \x_\star)$ is the likelihood---the probability of observing $(X, Y, \y_\text{exp})$ given a proposal $\x_\star$---and $P(\x_\star)$ is the \emph{prior}, which encapsulates initial knowledge of $\x_\star$.

In this study, we place informative priors on the entropy deposition parameter $p$ using previous results at midrapidity.
Specifically, we set the priors on $p$ equal to the posterior distributions determined by the Bayesian analysis in Ref.~\cite{Bernhard:2016tnd} which was calibrated to fit charged particle yields, mean transverse momenta, and flows at midrapidity.
For all remaining parameters, we assign a flat prior which is constant within the design range and zero outside.

We assume a Gaussian form for the likelihood function,
\begin{align}
  \begin{aligned}
  P &= P(X, Y, \y_\text{exp}|\x_\star) \\
    &= P(X, Z, \z_\text{exp}|\x_\star) \\
    &\propto\exp\biggl\{-\frac{1}{2}(\z_\star - \z_\text{exp})\trans \Sigma_z^{-1}(\z_\star - \z_\text{exp})\biggr\},
  \end{aligned}
  \label{eq:likelihood}
\end{align}
which is evaluated using the emulated principal components $\z_\star$ and transformed experimental data $\z_\text{exp}$.
Here $\Sigma_z$ is the covariance matrix for the principal components, which accounts for the various sources of uncertainty. We used a covariance matrix in the principal component space proportional to the identity matrix $\Sigma_z = \sigma I$, which corresponds to $5\%$, $10\%$, and $20\%$ relative error on the total variance of $\dndypP$,\, $\dndyPP$ and rms $a_1$.
This procedure effectively gives a larger weight to the p+Pb dataset, as it is more sensitive to the asymmetry parameters of the models, and it also emphasizes fitting single-particle observables over two-particle correlation observables.

Finally, the total log-likelihood of the model given by the p+Pb and Pb+Pb datasets is
\begin{align}\label{naive-formula}
\nonumber  \ln P &= \ln P_\text{pPb, $dN/d\eta$} + \ln P_\text{PbPb, $dN/d\eta$} \\
  &+  \ln P_\text{PbPb, rms $a_1$} + \ln P_{\text{priori}},
\end{align}
where $P_\text{piori}$ is the initial prior distribution.
The posterior is finally constructed by sampling the distribution using an affine-invariant MCMC sampler \cite{emcee}.
In each MCMC step, the Gaussian process emulators first predict the principal components of the model outputs, the likelihood is then computed from Eq.~\eqref{eq:likelihood}, and the posterior probability is calculated from Bayes' theorem~\eqref{bayes}.
We use $\mathcal O(10^5)$ burn-in steps and $\mathcal O(10^6)$ production steps to generate the posterior distribution.

A model evaluation measure known as a Bayes factor can then be used to compare the performance of the relative- and absolute-skewness models.
It is defined as the ratio of the likelihood functions, integrated over each model's parameter space,
\begin{eqnarray}
	K = \frac{\int P(\textrm{Exp}|\textrm{Model I}, \vec{p}) d\vec{p}}{\int P(\textrm{Exp}|\textrm{Model II}, \vec{p}) d\vec{p}}\,.
\end{eqnarray}
The interpretation of the scale of $K$ is listed in Table \ref{tab:Kfactor}. \cite{Jeffreys:1961}.
\begin{table}[t]
  \caption{Interpretation of the scale of $K$}
  \begin{ruledtabular}
    \begin{tabular}{ll}
      K & Strength of evidence (supports model I) \\
      \paddedhline
      $ < 10^{1/2}$	& negative (supports model II)	\\
	  $10^0$ -- $10^{1/2}$	& barely worth mentioning	\\
	  $10^{1/2}$ -- $10^1$	& substantial	\\
	  $10^1$ -- $10^{3/2}$	& strong	\\
	  $10^{3/2}$ -- $10^2$	& very strong	\\
	  $>10^2$	& decisive	\\
    \end{tabular}
  \end{ruledtabular}
  \label{tab:Kfactor}
\end{table}
With $K<1$, the experimental data supports model II; while increasing $K$ above $1$, model I is supported with increasing strength of evidence.

\section{Calibration results}
\subsection{Posterior probability distribution}
Fig.~\ref{fig:posterior} presents the Bayesian posterior probability distributions for the relative- and absolute-skewness models (blue lower- and red upper-triangular matrices respectively).
Diagonal panels show the marginal posterior distribution of individual model parameters (all other parameters integrated out), while off-diagonal panels show the joint distribution for pairs of model parameters, reflecting their correlations.

The posterior distributions contain a wealth of information; here we summarize a few key observations:
\begin{itemize}[itemsep=0pt, leftmargin=2\parindent]
  \item Both models prefer the entropy deposition parameter $p$ close to $0$, consistent within the range of the prior distribution extracted from \cite{Bernhard:2016tnd}.
  \item The p+p multiplicity fluctuation parameter is well constrained and distributed about $k=2.0$ for both models.
    These $k$ values are also consistent with the range of the previous estimates obtained from fits to p+p, p+Pb, and Pb+Pb multiplicity distributions at midrapidity \cite{Moreland:2014oya}.
  \item The relative-skewness model prefers a larger nucleon width than the absolute-skew model. 
  For future studies, one may also use more granular protons with subnucleonic structure instead of Gaussian protons.
  \item The calibrated relative-skewness model exhibits almost zero shift about the mean and large skewness, while the absolute-skewness model prefers a shift close to the center-of-mass rapidity and a moderate skewness. 
  The two different parametrizations thus prefer qualitatively different mechanisms for asymmetric entropy deposition.
\end{itemize}

\begin{figure*}
  \includegraphics{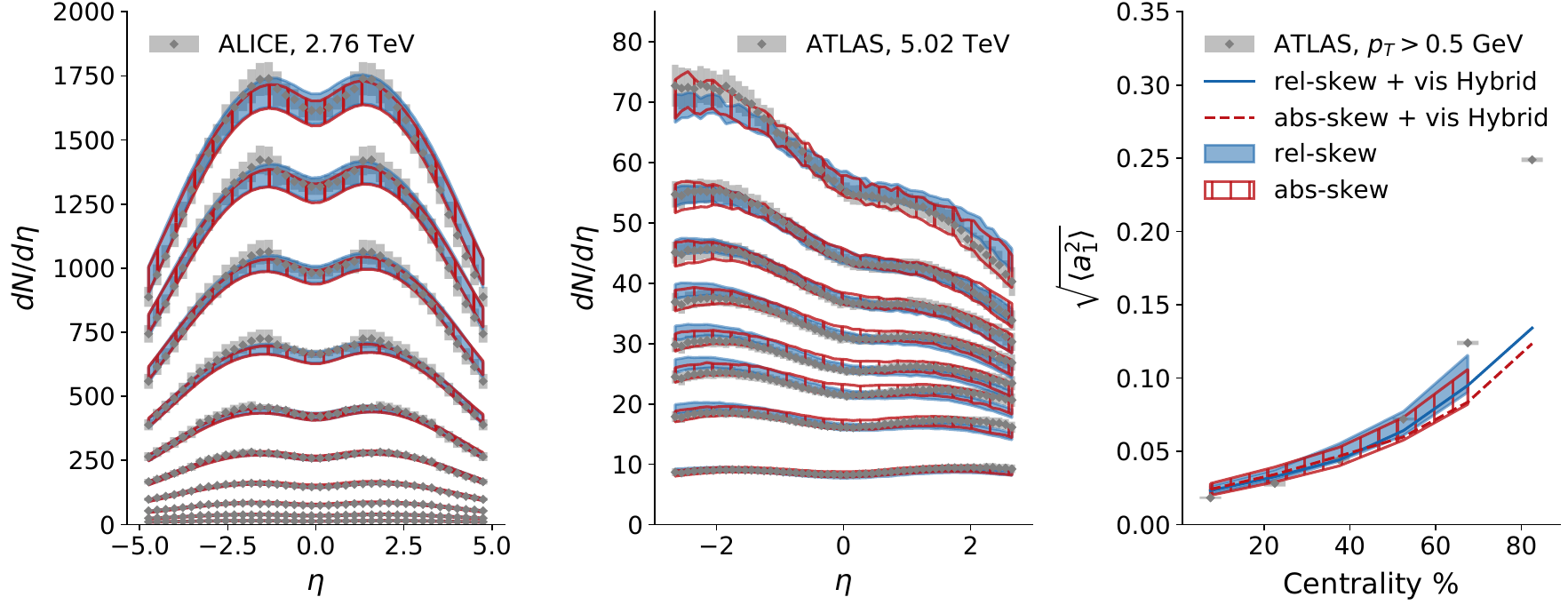}
  \caption{
  Left and Middle: Centrality and pseudorapidity dependence of the charged particle pseudorapidity density $\dndyPP$ at $\sqrts=2.76$~TeV and $\dndypP$ at $\sqrts=5.02$~TeV.
  Bands represents model emulator calculations $\dnchdy$ from the Bayesian posterior.
  Symbols are experimental data from ALICE \cite{Abbas:2013bpa,ALICE:2015kda} and ATLAS \cite{Aad:2015zza}.
  Rapidity cuts on model centrality selection are matched to experiment.
  Right: 
  The root-mean-square of the Legendre expansion coefficient $a_1$ estimated from two-particle pseudorapidity correlations plotted as a function of collision centrality.
 Bands represents model emulator calculations $\dnchdy$ from the Bayesian posterior, while lines are results from full event-by-event viscous hybrid model simulations using selected parameters from the Bayesian posterior.
  Symbols with errors are experimental data from ATLAS \cite{SoorajRadhakrishnanfortheATLAS:2015eqq}.  
  }
  \label{fig:post_obs}
\end{figure*}

\subsection{Calibrated observables}

To visualize the performance of the two models, we show in Fig.~\ref{fig:post_obs} the resulting observables calculated from each model's calibrated emulators.
The bands are centered around the mean prediction, and their spread denotes $\pm 2$ standard deviations.
Both calibrated models are able to simultaneously describe $\dnchdy$ for the two collision systems as functions of rapidity and centrality, illustrating the flexibility of the generating function approach.

The results for rms $a_1$ are compared to preliminary data from ATLAS \cite{ATLAS:2015kla} in Fig.~\ref{fig:post_obs}.
Both models capture the increasing trend of rms $a_1$ as function of centrality.
Hybrid model calculations agree with experimental measurements within $20\%$ for $0$--$50\%$ centralities ($N_{\textrm{part}} \gtrsim 75$) but underestimate the data at more peripheral centralities.
We notice that in \cite{ATLAS:2015kla}, \mbox{HIJING} calculations reproduce rms $a_1$ for $N_{\textrm{part}} \lesssim 80$ but overpredict the signal at larger $N_{\textrm{part}}$.
This suggests that hydrodynamic calculations and microscopic models are complementary in understanding longitudinal fluctuations.

Averaging the likelihood function over an ensemble of posterior parameter sets for each model gives the model likelihood, from which the Bayes factor is calculated,
\begin{eqnarray}
K = \frac{\text{Relative-skewness model}}{\text{Absolute-skewness model}} = 2.5 \pm 0.2. 
\end{eqnarray}
This value is too close to unity to make a decisive statement regarding the preference of one model of the other \cite{Jeffreys:1961}.
Indeed, the absolute-skewness model is slightly better in capturing the asymmetries in p+Pb collisions; while the relative-skewness model exhibits a larger curvature for $\sqrt{\langle a_{1}^2 \rangle}$, closer to experiment.

In summary, both models describe the p+Pb and Pb+Pb charged-particle pseudorapidity densities in all centrality bins to 10\% accuracy.
It also describes the rms $a_1$ from central to mid-central Pb+Pb collisions.
Both models fail to describe the rms $a_1$ in peripheral collisions which suggests that additional sources of fluctuation are needed in addition to nuclear thickness function fluctuations.
Relevant sources could include initial dynamical fluctuations such as string fragmentation, subnucleonic fluctuations, and finite-particle effects.

\section{Predictions on novel observables}
Both relative- and absolute-skewness models provide comparable descriptions of multiplicity observables in p+Pb and Pb+Pb collisions.
In this section, we proceed to investigate whether they can describe azimuthally sensitive observables such as flows, event-plane decorrelations and symmetric cumulants.
Here we use selected initial condition parameters around the peaks of the posterior distributions for each model (Table \ref{tab:chosen_parameters}) and perform viscous 3+1D hydrodynamics evolution with UrQMD as an afterburner.
These observables provide a nontrivial test of the proposed model as they resolve azimuthal correlations which have not been included in the calibration process.

\begin{table}[t]
  \caption{Selected high-probability parameter sets}
  \begin{ruledtabular}
    \begin{tabular}{lll}
      Parameter & rel-skew	& abs-skew \\
      \paddedhline
      $N_{\textrm{Pb+Pb}}^\dagger$   & 150.0     & 154.0  \\
      $p$	    & 0.0      & 0.0  \\
      $k$	    & 2.0     & 2.0  \\
      $w$	    & 0.59     & 0.42  \\
      $\mu_0$   & 0.0     & 0.75  \\
      $\sigma_0$ & 2.9    & 2.9  \\
   	  $\gamma_0$ & 7.3		& 1.0	\\
      $J$	     & 0.75 & 0.75	\\
    \end{tabular}
  \end{ruledtabular}
  \raggedright{$\dagger$ Normalization tuned with ideal hydro is reduced when using viscous hydro.}
  \label{tab:chosen_parameters}
\end{table}

\subsection{Anisotropic flows} 
\begin{figure}
  \includegraphics{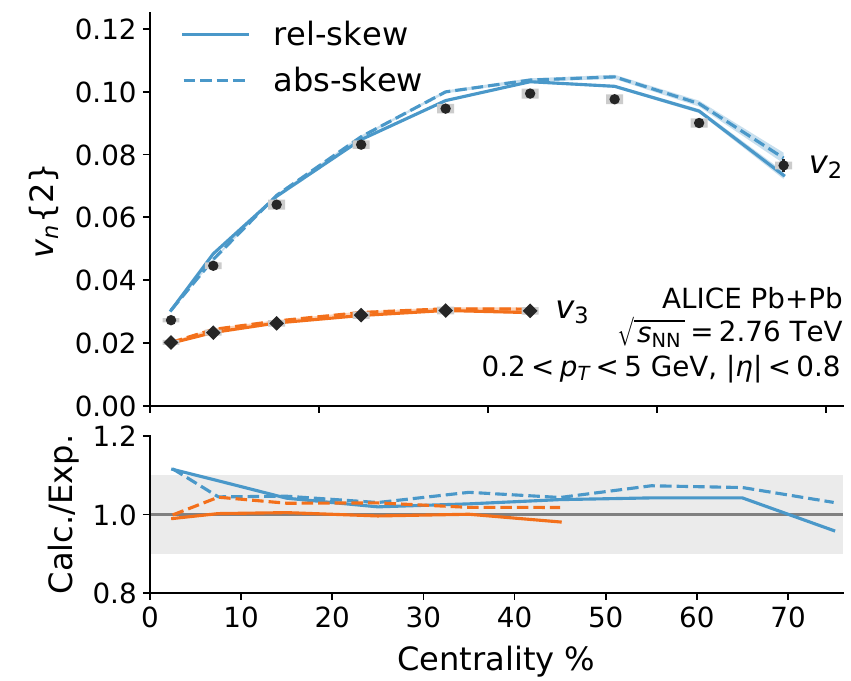}
  \caption{Elliptic and triangular flow cumulants $v_2\{2\}$ and $v_3\{2\}$ as a function of centrality calculated from 3+1D hybrid model simulations using constant specific shear viscosity $\eta/s=0.17$ and $0.19$ for relative- and absolute-skewness models respectively, zero bulk viscosity $\zeta/s=0$ and hydro-to-micro switching temperature $T_\text{sw}=154$~MeV.
  The initial condition parameters are selected from the Bayesian posterior.}
  \label{fig:vn_cen}
\end{figure}

As a preliminary check, we first verify that previous results for the elliptic and triangular flow harmonics $v_2\{2\}$ and $v_3\{2\}$ obtained using \trento\ initial conditions at midrapidity \cite{Bernhard:2016tnd} are indeed recovered by the rapidity-dependent model extension.
Fig.~\ref{fig:vn_cen} shows the centrality dependence of $p_T$-integrated flow for ${0.2  < p_T < 5.0}$~GeV and $|\eta| < 0.8$ calculated from the \emph{three-dimensional} hybrid model compared to \mbox{ALICE} measurements \cite{Adam:2016izf} using the  $Q$-cumulant method \cite{Bilandzic:2010jr}.
The 3+1D hydrodynamics code used in this study only partially implements bulk viscous corrections and thus is not yet suitable for quantitative calculations involving finite bulk viscosity.
We therefore assert a QGP specific bulk viscosity $\zeta/s = 0$ which precludes direct comparison with the boost-invariant VISH2+1 hydrodynamics code \cite{Song:2007ux, Shen:2014vra, Bernhard:2016tnd} and the corresponding shear and bulk viscosities determined by the previous Bayesian analysis \cite{Bernhard:2016tnd}.
For the QGP specific shear viscosity, we choose constant QGP $\eta/s = 0.17$ and $0.19$ for relative- and absolute-skewness models respectively, which provide good descriptions of the data \cite{Gale:2012rq, Niemi:2015qia}, although it is not a systematic best fit.
The resulting $v_2\{2\}$ and $v_3\{2\}$ agree with experimental data within $10\%$ and verify that the generating function rapidity extension recovers previous \trento\ initial condition results at midrapidity.

We now proceed to calculate the pseudorapidity-dependent flows which provide a sensitive handle on the QGP transverse structure at different rapidity values.
The ALICE collaboration has measured $v_n\{2\}(\eta)$ and $v_2\{4\}(\eta)$ within the wide pseudorapidity interval $-3.5 < \eta < 5.0$ and extrapolated to zero $p_T$ \cite{Adam:2016ows}. 
This extrapolation reduces integrated flow relative to measurements with a nonzero $p_T$ cut because it averages over low-$p_T$ particles which generally have less flow.
The same behavior occurs in hydrodynamic models, although models which mispredict mean $p_T$ also mispredict the corresponding change in flow produced by introducing a $p_T$ cut.
The hybrid model used in this study omits bulk viscous corrections and thus overpredicts mean $p_T$.
This means it cannot describe hydrodynamic flow measurements with different $p_T$ cuts using a single value of $\eta/s$.
To circumvent this issue, we use $\eta/s=0.25$--$0.28$ when comparing to ALICE measurements that are extrapolated to zero $p_T$. 
Future implementation of realistic bulk viscous corrections would eliminate such fine tuning.

\begin{figure*}
  \includegraphics{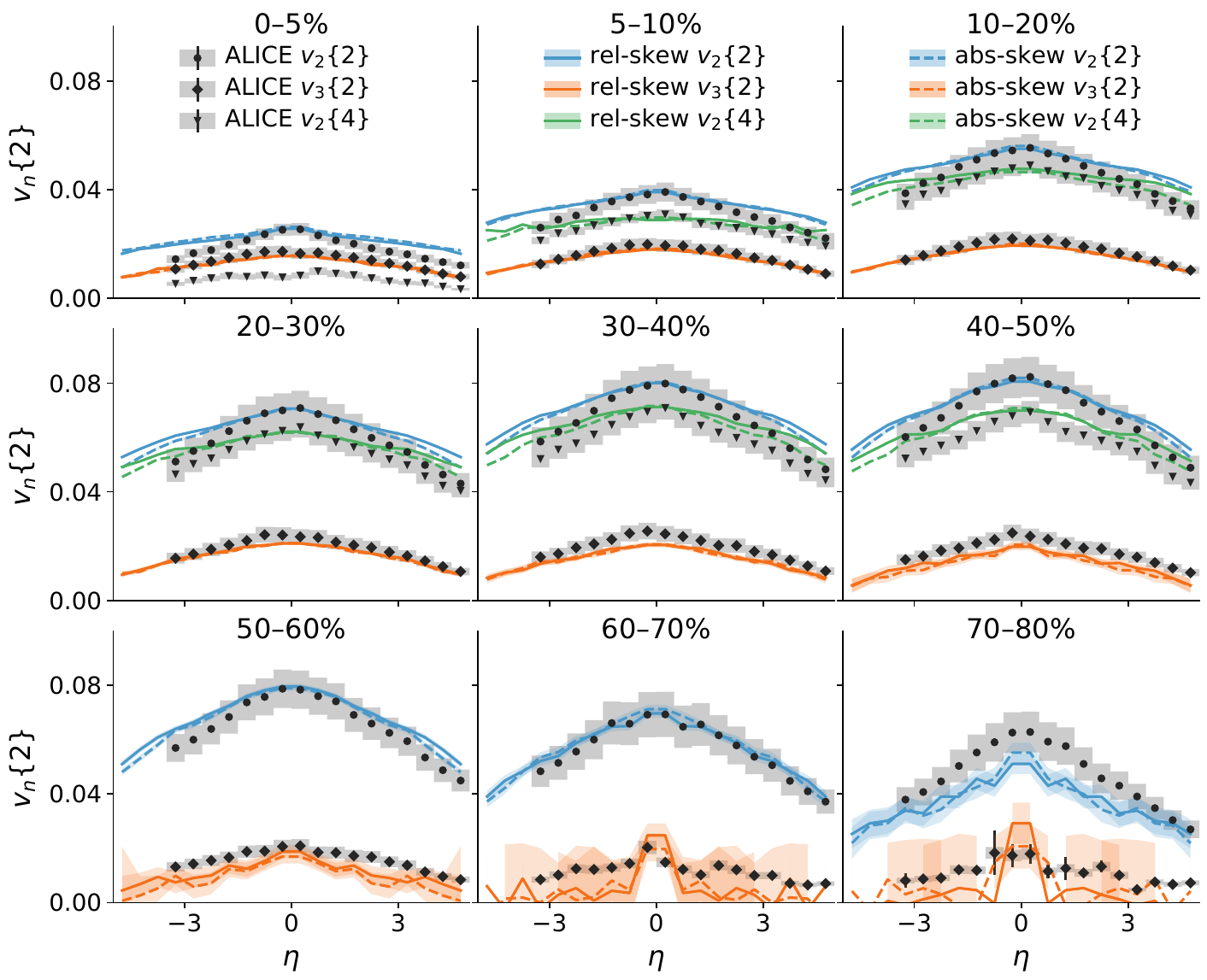}
  \caption{Pseudorapidity dependence of anisotropic flow coefficients $v_2\{2\}$, $v_2\{4\}$ and $v_3\{2\}$ (blue, green and orange lines) calculated from the hybrid model with constant specific shear viscosity $\eta/s=0.25$ and $0.28$ for relative- and absolute-skewness models respectively, compared to data from ALICE (symbols) with $p_T > 0$~GeV (extrapolated) in different centrality bins. 
  Theory bands indicate $1\sigma$ statistical error, while experimental bands/bars denote $1\sigma$ systematic and statistical errors respectively.}
  \label{fig:vn_eta}
\end{figure*}

The pseudorapidity-dependent flows are estimated using the cumulant approach \cite{Bilandzic:2010jr}, where particles of interests (POI) are correlated with reference particles.
The differential flow is then calculated via,
\begin{eqnarray}
v_n^\prime\{2\} = \frac{d_n\{2\}}{\sqrt{c_n\{2\}}},\\
v_n^\prime\{4\} = \frac{-d_n\{4\}}{\left(-c_n\{4\}\right)^{3/4}},
\end{eqnarray}
where $d_n\{2\}$, $d_n\{4\}$ is the two- and four-particle cumulants between the POI and reference particles and $c_n\{2\}$ and $c_2\{4\}$ are the cumulants among reference particles.
For POI with $\eta > 0$ ($\eta < 0$), the reference particles are restricted to $-0.8 <\eta < 0$ ($0 <\eta < 0.8$) to avoid autocorrelations.
The results are shown in the left panel of Fig.~\ref{fig:vn_eta} for nine centrality classes.
The correlation functions $d_n(\eta)$ and $c_n(\eta)$ are symmetrized since the event-averaged pseudorapidity-differential flow for the Pb+Pb system should be invariant with respect to the substitution $\eta \rightarrow -\eta$.

Both models predict $v_2\{2\}$, $v_2\{4\}$ and $v_3\{2\}$ that decrease from mid to forward/backward rapidity and produce a triangle shaped structure as measured by ALICE.
Incidentally, the absolute-skewness model agrees with experiment slightly better at large pseudorapidity.
It has been realized that the slope of $v_n(\eta)$ which produces this triangular shape is highly sensitive to the hadronic shear viscosity \cite{Denicol:2015bnf}, and thus Fig.~\ref{fig:vn_eta} corroborates that UrQMD provides a semi-quantitative description of hadronic viscosity below the QGP transition temperature.
However, for central to mid-central collisions, the slope of the decreasing $v_2$ as a function of pseudorapidity is underpredicted, resulting in a flatter $v_n(\eta)$ than the experiments.
The reason for this discrepancy may be complicated.
Apart from improving initial conditions, a realistic bulk viscosity and a temperature dependent specific shear viscosity should definitely affect the results.
Another reason could be the use of the QCD EoS and QGP transport coefficient $\eta/s$ in the limit of vanishing baryon chemical potential, which may not be a good approximation at large pseudorapidity even at LHC energies.

\begin{figure*}
  \begin{center}
  \includegraphics{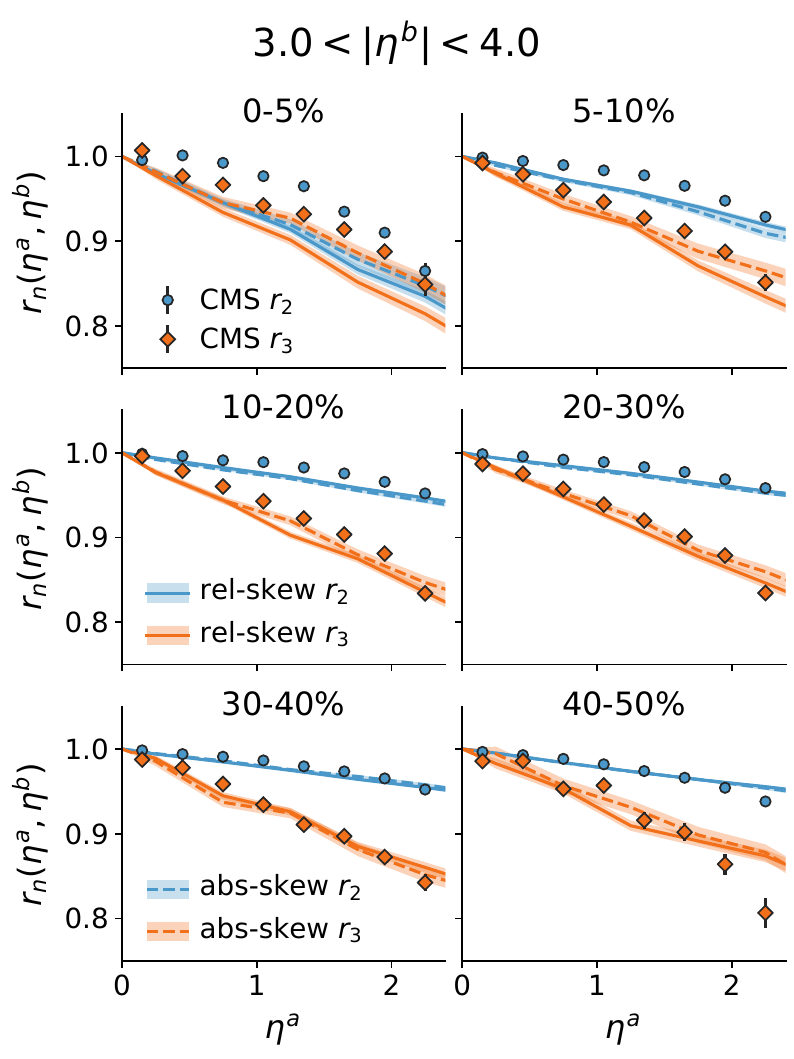}
  \quad\quad\quad
  \includegraphics{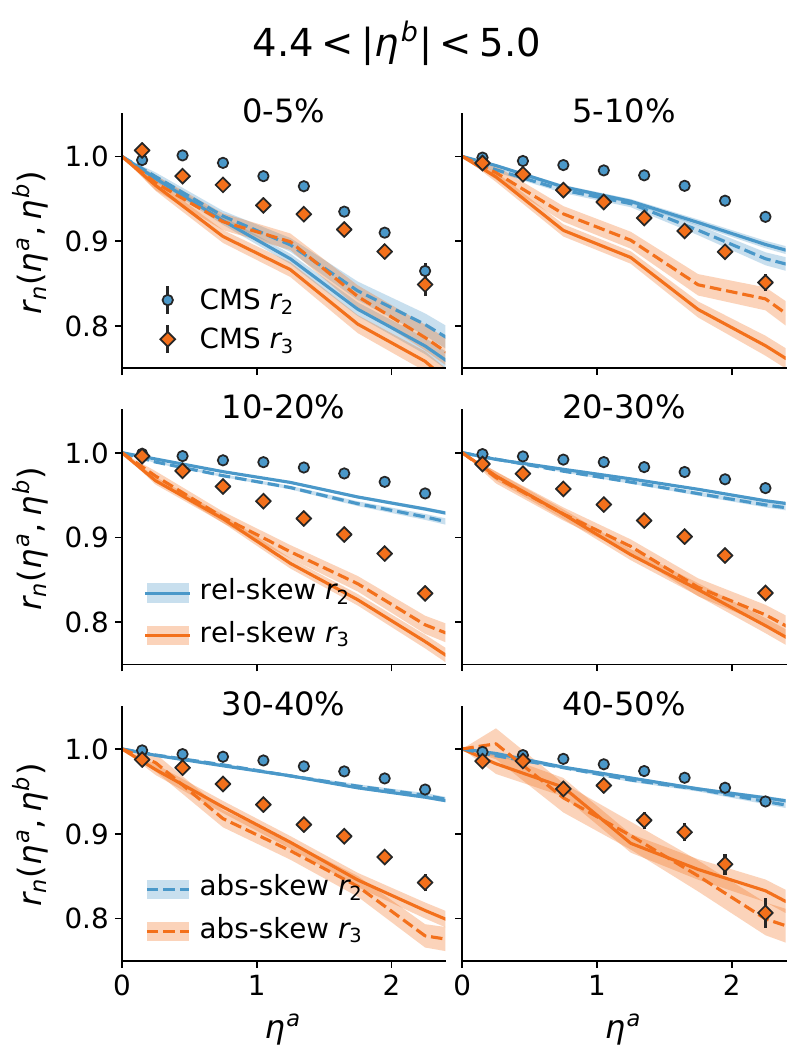}
  \end{center}
  \caption{Left: The event-plane decorrelation for $n=2,3$ in different centrality bins with the reference particles from $3.0<|\eta^b|<4.0$.
  Right: The same quantities as the left panel but with the reference particles from $4.4<|\eta^b|<5.0$. 
  Theory bands indicate $1\sigma$ statistical error, while experimental bands/bars denote $1\sigma$ systematic and statistical errors respectively.}
  \label{fig:epd}
\end{figure*}

\subsection{Event-plane decorrelation}

Next, we study the event-plane decorrelation as a function of pseudorapidity using the calibrated relative-skewness model. 
The event planes are defined by the angles
\begin{equation}
  \Psi_n^\text{EP} = \frac{\text{atan2}(\langle \sin n \phi \rangle, \langle\cos n \phi \rangle)}{n},
\end{equation}
where the average is performed over particles of interest.
In general, the angles $\Psi_n^\text{EP}$ may change as a function of pseudorapidity due to longitudinal initial state fluctuations and finite particle effects.
As a consequence, two event-plane angles constructed from sets of particles separated by a finite rapidity gap will decorrelate as the rapidity gap increases.
This effect is important as it affects not only the calculation of soft observables involving a finite pseudorapidity gap or a large pseudorapidity interval, but also the interpretation of hard probe observables where particles from a rare hard process are often correlated with reference particles from different pseudorapidity bins.
It has been studied in a number of previous works including a longitudinally torqued fireball model with fluctuating sources \cite{Bozek:2015bna}, AMPT calculations which studied its influence on flow observables \cite{Jia:2014ysa, Xiao:2012uw}, as well as coarse-grained AMPT initial conditions that were embedded in 3+1D ideal hydrodynamic simulations \cite{Pang:2015zrq}.

The decorrelations receive contributions from both random fluctuations during the evolution process and the systemic twist of the participant plane arising from initial longitudinal fluctuations \cite{Bozek:2015bna}.
In the present parametric initial condition model, the participant plane twist arises naturally from local longitudinal fluctuations.
The transverse geometry at forward (backward) space-time rapidity is dominated by the projectile (target) participant density.
As a result, the participant plane gradually interpolates between the projectile and target densities, leading to a systemic twist in the beam direction.
The time evolution also contributes to decorrelation among the event-planes.
For example, early- and late-stage dynamics introduce additional fluctuations that partially randomize event-plane orientations.
Stochastic contributions from pre-equilibrium dynamics are neglected in the present study, but fluctuations in the hadronic phase are naturally accounted for by the UrQMD transport model.

The CMS collaboration has measured the event-plane decorrelations in Pb+Pb collisions using the $\eta$-dependent factorization ratio $r_n(\eta^a, \eta^b)$ \cite{Khachatryan:2015oea}, defined as
\begin{align}
  r_n(\eta^a, \eta^b) &= \frac{V_{n\Delta}(-\eta^a, \eta^b)}{V_{n\Delta}(\eta^a, \eta^b)}, \\
  V_{n\Delta}(\eta^a, \eta^b) &= \langle\langle \cos(n\Delta\phi) \rangle\rangle,
\end{align}
where the double average means averaging over particles in each event and then averaging over all events in a given centrality class. 
The use of three $\eta$-bins ($\pm \eta^a$ and $\eta^b$) reduces short range correlations.
The ratio $r_n(\eta^a, \eta^b)$ reflects the fluctuation of event-plane angles separated by $\eta^a+\eta^b$ relative to the fluctuation of angles separated by  $|\eta^a-\eta^b|$ \cite{Khachatryan:2015oea}.

We compare our calculation to the CMS measurements with both $3.0 < \eta^b < 4.0$ and $4.4 < \eta^b < 5.0$ and momentum cuts $p_T^b > 0$~GeV and ${0.3 < p_T^a < 3.0}$~GeV.
The $\eta$-dependent factorization ratios $r_2$ and $r_3$ for six centrality classes and different $\eta^b$ cuts are shown in Fig.~\ref{fig:epd}.
Both models predict a prominent $n=2, 3$ event-plane decorrelation in central collisions and decreases with increasing centrality.
For midcentral collisions, the nuclear geometry largely defines the $n=2$ participant plane---fluctuations and twisting are perturbations around this predominant direction---and hence $r=2$ decorrelation is reduced.
On the other hand, the $n=3$ event-plane receives little contribution from the nuclear geometry but is dominated mostly by fluctuations; it therefore has a similar slope over all six centralities.
In central collisions, the contribution from nuclear geometry is overwhelmed by fluctuations leading to similar $n=2$ and $n=3$ decorrelations.
The calculations describe the observed $n=2,3$ event-plane decorrelations with $3.0 < \eta^b < 4.0$ very well except the most central $0$--$5\%$ centrality, but systematically overpredict the magnitude of the decorrelations with $4.4 < \eta^b < 5.0$, especially for $0$--$10\%$ central collisions.
The reason is that the model, by construction, extends well-developed mid-rapidity initial conditions to finite pseudorapidity. 
Even though it is calibrated to multiplicity observables, it gradually loses its predictive power for fine-structure flow observables when moving far away from mid-rapidity.
Specifically, the model predicts decorrelations between the event-planes that are stronger for larger $\eta^b$ bins, while the experiment sees that the magnitude of decorrelation saturates when moving from $3.0<\eta^b<4.0$ to $4.4<\eta^b<5.0$.
Future improvements to the model at large pseudoraidity are clearly needed.
Nevertheless, the model's explanation of the event-plane decorrelations for $3.0 < \eta^b < 4.0$ remains nontrivial.
Both models were both calibrated with $\dnchdy$ and rms $a_1$ data.
These multiplicity observables do not constrain the transverse structure of the event at different pseudorapidities, and hence reproducing $r_2$ and $r_3$ means the calibrated initial condition models not only reproduce global longitudinal entropy deposition and fluctuations, but also capture features of the longitudinal dependence of transverse geometry within $|\eta| \lesssim 4$.

\begin{figure*}
  \includegraphics{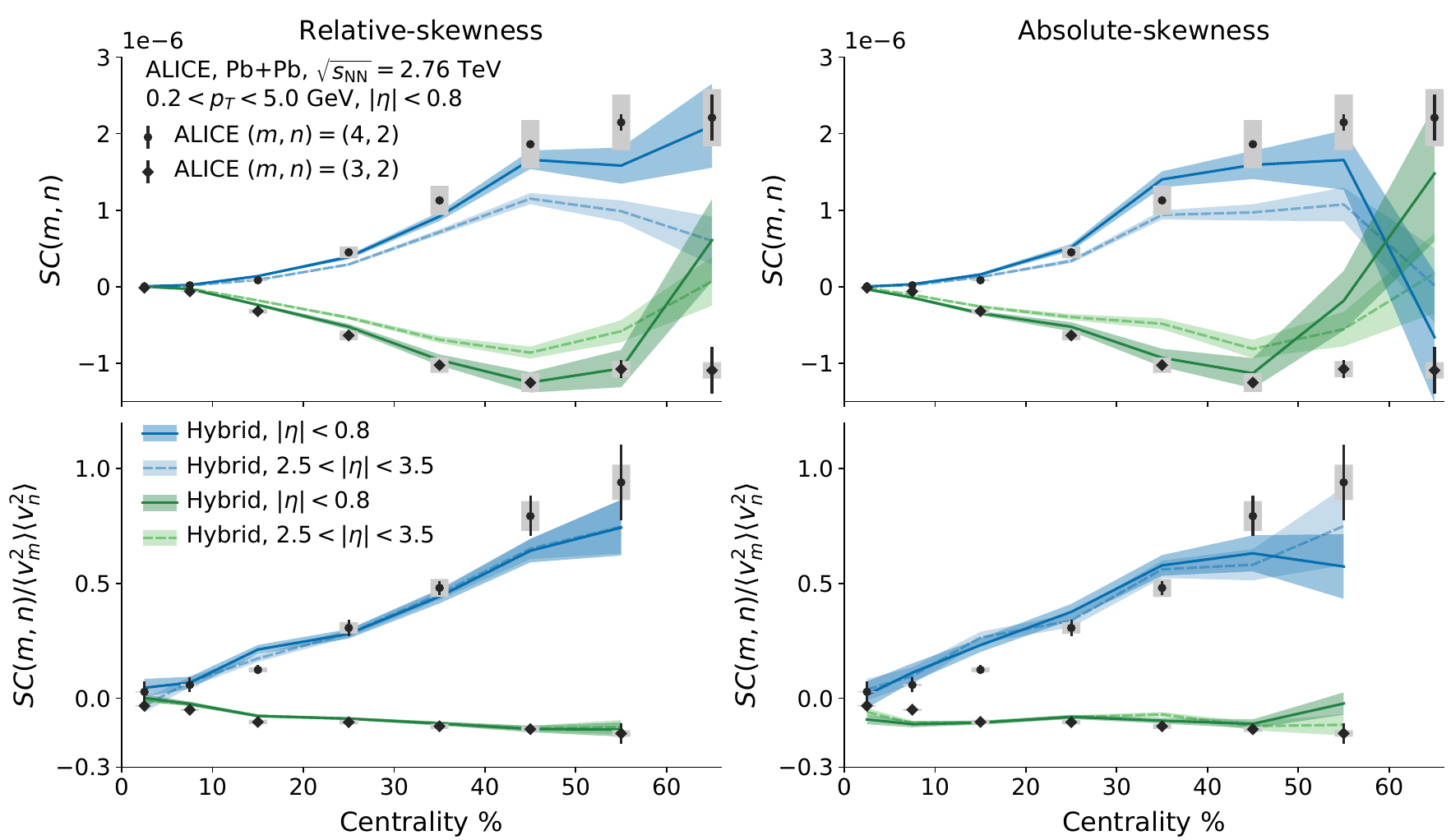}
  \caption{Top: calculated $SC(4,2)$ and $SC(3,2)$ as functions of centrality compared to ALICE measurements, with the same 3D hybrid model set-up as used for Fig. \ref{fig:vn_cen}.
  We conduct calculations in two kinematic ranges: $|\eta|<0.8$ is the pseudorapidity cut used by current the ALICE measurement and $2.5<|\eta|<3.5$ is our prediction for the symmetric cumulants away from mid-rapidity in the Pb+Pb system.
  Bottom: $SC(m,n)$ normalized by $\langle v_m^2\rangle\langle v_n^2\rangle$ for two two kinematic ranges.
   }
  \label{fig:smn} 
\end{figure*}

\subsection{Flow correlations}

Correlations between different anisotropic flow harmonics can be used to further constrain the initial state geometry \cite{Niemi:2012aj}.
Experimentally, these correlations can be quantified using either event shape engineering \cite{Schukraft:2012ah, Aad:2015lwa} or the symmetric cumulants $SC(m,n)$ \cite{Bilandzic:2013kga}. 
Here we focus on the symmetric cumulants which are defined as,
\begin{align}
SC(m, n) &= \langle\langle \cos(m\phi_1+n\phi_2-m\phi_3-n\phi_4)\rangle\rangle \nonumber \\
\nonumber &- \langle\langle\cos[m(\phi_1-\phi_2)]\rangle\rangle\langle\langle\cos[n(\phi_1-\phi_2)]\rangle\rangle \label{eq:scmn}\\
&= \langle v_m^2 v_n^2 \rangle - \langle v_m^2\rangle\langle v_n^2\rangle.
\end{align}
The centrality dependences of $SC(4,2)$ and $ SC(3,2)$ at midrapidity have recently been measured by ALICE \cite{ALICE:2016kpq}. 
A positive value of $SC(m,n)$ means that a large $v_m$ is more likely to be observed with a large $v_n$, while for negative values of $SC(m,n)$, a large $v_m$ favors small $v_n$.
The symmetric cumulants $SC(m,n)$ are nearly insensitive to nonflow effects while remaining sensitive to collective effects, initial geometry fluctuations $\langle \varepsilon_m^2 \varepsilon_n^2 \rangle - \langle \varepsilon_m^2 \rangle \langle \varepsilon_n^2 \rangle$ and the QGP specific shear viscosity \cite{ALICE:2016kpq, Zhu:2016puf}.
To remove its dependence on the magnitudes of $\langle v_m^2\rangle$ and $\langle v_n^2\rangle$, we also calculate the normalized symmetric cumulants
\begin{equation}
  NSC(m,n) = SC(m,n)/\langle v_m^2\rangle\langle v_n^2\rangle.
\end{equation}
Here we use this tool to not only study the flow correlations at midrapidity, but also reveal its pseudorapidity dependence.

Fig.~\ref{fig:smn} shows the calculated symmetric cumulants compared to ALICE measurements using the relative- and absolute-skewness models with the same transport coefficients as in Fig.~\ref{fig:vn_cen}. 
We use the same centrality bins as ALICE experiments and the centrality averaged symmetric cumulants are performed with a multiplicity weight as discussed in \cite{Gardim:2016nrr}.
We first calculate $SC(4,2)$ and $SC(3,2)$ at midrapidity $|\eta|<0.8$ (solid lines) to match the rapidity cuts of the ALICE measurement.
The negative $SC(3,2)$ is a result of initial eccentricity correlations, while the large positive $SC(4,2)$ is produced by nonlinear correlations between $v_2$ and $v_4$ during the medium evolution \cite{Giacalone:2016afq, Qian:2016fpi, Bhalerao:2014xra, Zhou:2015eya}.
The resulting symmetric and normalized symmetric cumulants agree with the data quite well and support previous constraints on the QGP initial conditions at midrapidity \cite{Bernhard:2016tnd}. 

Next, we shift our attention away from midrapidity and predict the symmetric (normalized symmetric) cumulants in the rapidity interval $2.5 < |\eta| < 3.5$ (dashed lines) which has not been measured.
In this calculation, we take two reference particles from $|\eta| < 0.8$ and two POI from $2.5 < |\eta| < 3.5$ and calculate
\begin{align}
SC^\prime(m, n) &= \langle\langle \cos(m\phi_1+n\phi_2-m\phi_3^\prime-n\phi_4^\prime)\rangle\rangle \\
\nonumber &- \langle\langle\cos[m(\phi_1-\phi_2^\prime)]\rangle\rangle\langle\langle\cos[n(\phi_1-\phi_2^\prime)]\rangle\rangle, \label{eq:scmn}
\end{align}
where the primed symbols represents the azimuthal angle of POI.
Magnitudes of both $SC^\prime(4, 2)$ and $SC^\prime(3, 2)$ are significantly suppressed at forward/backward rapidities, in accordance with the behavior of $v_2\{2\}(\eta)$ and $v_3\{2\}(\eta)$ as presented in the text in Fig.~\ref{fig:vn_eta}.
However the normalized symmetric cumulant $NSC^\prime(4,2)$ and $NSC^\prime(3,2)$ are consistent within uncertainty bands with different pseudorapidity cuts.
The unchanged normalized symmetric cumulant as function of psuedorapidity is a prediction of both relative- and absolute-skewness models and future comparison with available data should have strong constraining power to our approach to the three-dimensional initial conditions.

\section{Conclusion}

In summary, we have proposed a new method to extend arbitrary initial condition models defined at midrapidity to forward and backward pseudorapidity.
The method describes initial entropy deposition as a purely local function of nuclear participant densities, with the longitudinal profile reconstructed from generating-function cumulants.
The first three cumulants of the distribution (mean, standard deviation, and skewness) are included.
We set the mean proportional to the center-of-mass rapidity of local nuclear participant densities, and parametrize the standard deviation using a constant rapidity width.
Two models for the distribution's skewness are investigated: one where the skewness is proportional to the relative nuclear thickness difference, and one where it is proportional to the absolute difference.

We apply the method to extend the parametric \trento\ initial condition model which has been previously used to constrain QGP initial conditions and medium properties at midrapidity. 
The resulting three-dimensional models are then calibrated using Bayesian parameter estimation to fit p+Pb and Pb+Pb charged particle pseudorapidity densities $\dnchdy$ and the root-mean-square of the two particle pseudo-rapidity correlation's Legendre decomposition coefficient $a_1$ at the LHC.
After the calibration, both models have comparable qualities of describing $\dnchdy$ and rms $a_1$.

Using the calibrated relative- and absolute-skewness initial condition models, we study pseudorapidity-dependent anisotropic flows, event-plane decorrelations and flow correlations in Pb+Pb collisions.
The model nicely describes integrated flows $v_2$ and $v_3$ at midrapidity as well as the pseudorapidity dependence of differential flow for different centrality classes.
The elliptic and triangular event-plane decorrelations with $3.0 < |\eta^b| < 4.0$ are well explained except for the most central collisions, but both models overpredict the decorrelations with the reference particles $4.4 < |\eta^b| < 5.0$.
This comparison sets the range of validity for flow/event-plane calculations using the two models.
Moreover, both models give a satisfactory description of flow correlation $SC(3,2)$ and $SC(4,2)$ at midrapidity, which can be used to predict their values at forward/backward pseudorapidity where their values have not yet been measured.

The present work expands upon previous efforts to parametrize and constrain local initial condition properties using global final-state observables.
We show that these local properties are overconstrained by data and can be reverse engineered using systematic model-to-data comparison with quantitative uncertainty.
The resulting knowledge can then be used to provide direct feedback for first-principle calculations of the QGP initial conditions.
The present analysis would benefit from a number of future improvements.
It would be interesting to add subnucleonic structure to the nuclear thickness functions in order to examine its effect on longitudinal rapidity fluctuations.
We leave these refinements to future studies.

\begin{acknowledgments}
  This research was completed using $10^6$ CPU hours provided by the Open Science Grid \cite{Pordes:2007zzb,Sfiligoi:2010zz}, which is supported by the National Science Foundation and the U.S.\ Department of Energy's Office of Science.
SAB and WK  are supported by the U.S. Department of Energy Grant no. DE-FG02-05ER41367, JEB is supported by NSF grant ACI-1550225 and JSM by the DOE/NNSA Stockpile Stewardship Graduate Fellow- ship under Grant no. DE-FC52-08NA28752.
  We thank Jianyong Jia for providing ATLAS preliminary data and Jacquelyn Noronha-Hostler for useful discussion.
\end{acknowledgments}
\appendix*

\section{Hybrid model simulation}\label{app}

The 3+1D relativistic viscous hydrodynamics code \mbox{vHLLE} \cite{Karpenko:2013wva} is used for the QGP medium evolution. 
The equation-of-state (EoS) is obtained by interpolating a state-of-the-art lattice-QCD EoS \cite{Bazavov:2014pvz} at high temperature with vanishing baryon density and a hadron resonance gas EoS at low temperature.
We use a switching energy density $\varepsilon_s = 0.322$~GeV/fm$^3$ ($T_s\sim0.154$~GeV) at which the hydrodynamic description is switched to the UrQMD transport description. 
The switching temperature $T_s$ is the same as the EoS pseudo-critical temperature $T_c = 0.154$~GeV. 
The hydrodynamic transport coefficients are given by:
\begin{align}
  (\eta/s)(T>T_s)  &=  \text{0.17--0.28}, \\
  (\zeta/s)(T>T_s) &=  0.0.
\end{align}
For simplicity, there is no bulk viscosity and the shear viscosity to entropy ratio is a constant.
Below $T_s$, the hydrodynamic energy density is particlized into hadrons and UrQMD takes over the time evolution of the hadronic system.
No additional inputs for the transport coefficients are needed.

\bibliography{trento3d} 

\end{document}